# FPGA based Agile Algorithm-On-Demand co-processor

R. Pradeep, S. Vinay, Sanjay Burman, Dr. V. Kamakoti
Department of Computer Science
Indian Institute of Technology, Madras
Chennai, India
kama@iitm.ernet.in

## Abstract

*With growing computational needs of many real-world applications, frequently changing specifications of standards, and the high design and NRE costs of ASICs, an algorithm-agile FPGA based co-processor has become a viable alternative. In this article, we report about the general design of an algorith-agile co-processor and the proof-of-concept implementation.*

## 1. Introduction

Co-processors were initially designed to reduce the computational overload on the host processors, thus scaling-up the latter's performance [1] [2]. Typically, co-processors, such as graphics co-processors, are application-specific. The computations they perform are predetermined.

In this paper, we provide a single chip solution for an agile co-processor that can execute any one of a set of functions on-demand by the host computer. The idea here is to give this co-processor additional agility by making it capable of executing many such computationally intensive functions.

## 2. Architecture

### 2.1. Co-processor architecture

This section gives an overview of the proposed architecture of the *FPGA based Agile Algorithm-On-Demand co-processor*. Our proposed architecture consists of three major blocks:

1. A memory
2. A PCI based microcontroller
3. A partially reconfigurable FPGA device.

A block diagram of the proposed architecture is shown in **Figure 1**. The entire system sits on a PCI card which can be fitted to a standard desktop computer / work-station / server. The system can be operated by issuing instructions to the microcontroller through the PCI.

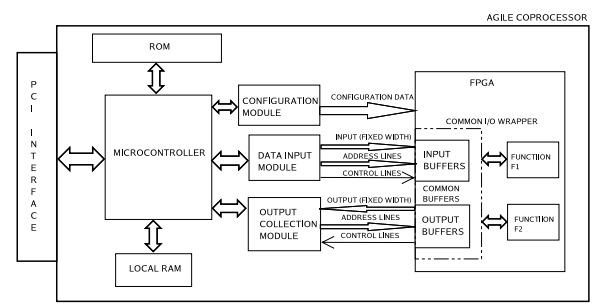

**Figure 1. A block diagram of the co-processor**

### 2.2. Memory

The compressed configuration bit-streams of the functions that are to be implemented on the FPGA are downloaded on to the ROM from the host PC. In addition, the ROM, contains records that holds the start address of each function's compressed configuration bit-stream on the ROM, its size and the input/output size of the functions. The compressed configuration bit-streams are loaded from one end of the ROM while the record table is populated from the other end of the ROM. The memory has an interface only with the microcontroller. The microcontroller utilises the records in the ROM to access the compressed bit-streams.

## 2.3. Microcontroller

The microcontroller is interfaced with the host PC(through the PCI), the ROM, the local RAM, the configuration module, the data input module and the output collection module. The microcontroller is responsible for accessing the compressed configuration bitstreams from the ROM and passing them to the configuration module. The microcontroller also takes inputs for the functions from the host through the PCI and stores them in the local RAM. It then passes these inputs to the data input module. Similarly the microcontroller collects the outputs from the functions through the data collection module and stores them in the local RAM before passing them back to the host PC.

The configuration module decompresses the compressed bit-stream window by window and passes the configuration bit-stream to the FPGA to configure it.

The data transfer to and from the FPGA takes place through the data input/output modules. Each data transfer is a multiple of the width of the interface bus as specified by the function record present in the ROM.

## 2.4. Partially reconfigurable FPGA chip

Parital reconfiguration of an FPGA is a method by which only a portion of the logic space on the FPGA is reconfigured. The configuration of the remaining portion of the FPGA is untouched and hence, the functions present in the untouched logic space may be accessed for execution. This parital reconfiguration of the FPGA facilitates the swap-in and swap-out of functions, from the FPGA, on-demand.

## 2.5. On-demand Algorithm

On-demand algorithms deals with the execution of a particular algorithm, from a bank of algorithms. When the host requests the execution of a particular algorithm, on some specified inputs, to the co-processor, the micro-controller is responsible for configuring the FPGA with that relevant configuration bit-stream if the function is not already present on the FPGA.

When the FPGA is programmed with the configuration bit stream of a particular algorithm, if occupies a certain number of *Frames*[1]. Thus, an algorithm's logic may occupy either a set of contiguous frames or a set of non-contiguous frames on the FPGA , as predetermined by the area constraints on the FPGA during the programming of the same. When a new algorithm is to be executed, the configuration bit stream of this algorithm may give rise to logic which may fit in the set of free frames, specified by the *Free Frame*

---

[1] Frames are a prespecified number of Logic Blocks and the relevant Switch Blocks of and FPGA

*List*. The micro-controller's mini OS maintains a Frames in the FPGA which are currently not used to realize any logic and are thus potentially programmable without any intervention to the functions currently being executed, called the Free Frame List.

However, if the Free Frame list is insufficient for the logic that the new algorithm's function is to realize, or is empty, some functions from the FPGA have to be erased so as to accommodate the logic necessary for the functions of the new algorithm whose execution has been requested by the host PC. The frames that are to be replaced is decided by the *Frame Replacement Policy* , which makes those frames that belong to the frequently least used Algorithm potential candidates for replacement. This set of frames may be decided by a look up in the Frame Replacement Table which gives an indication of the list of frames occupied by each algorithm present on the FPGA along with a time stamp specifying the last moment at which it was accessed. That algorithm which has the oldest time stamp provides extra frames for potential reconfiguration.

## 3. Proposed Proof-of-concept Implementation

The proof-of-concept implementation of the *FPGA based Agile Algorithm-On-Demand co-processor* is done using an Altera STRATIX PCI Development board [3] and a Xilinx Virtex-II FPGA which has partial reconfigurability [4].

## 4. Conclusion

This paper presents an algorithm agile co-processor. One of the interseting open problems is to explore advanced techniques for compression that can exploit the symmetry in the CLB architectures of FPGAs and also the partial reconfigurability of FPGAs.